\documentclass[12pt,preprint]{aastex}
\title{Evidence for solar frequency dependence on sunspot type}
\author{Charles S.~Baldner}
\affil{Department of Astronomy, Yale University, P.O. Box 208101, New Haven, CT, 06520-8101}
\author{Richard S.~Bogart}
\affil{Hansen Experimental Physics Laboratory, Stanford University, Stanford, CA 94305-4085}
\and
\author{Sarbani Basu}
\affil{Department of Astronomy, Yale University, P.O. Box 208101, New Haven, CT, 06520-8101}

\date{\today}

\begin{document}
\begin{abstract}
High degree solar mode frequencies as measured by ring diagrams are known to change in the 
presence of the strong magnetic fields found in active regions. We examine these 
changes in frequency for a large sample of active regions analyzed with data from 
the Michelson Doppler Imager (MDI) onboard the SoHO spacecraft, spanning most of 
solar cycle 23. We confirm that the frequencies increase with increasing magnetic 
field strength, and that this dependence is generally linear. We find that the 
dependence is slightly but significantly different for active regions with different 
sunspot types.
\end{abstract}

\section{Introduction}
Active regions are complex, dynamic objects and are prominent manifestations of
solar activity.  Understanding the nature, origin, and development of these regions
is one of the most important outstanding questions in solar physics.  Helioseismology
provides us with a means to probe the layers of the Sun below the surface of active
regions, and has the potential to yield important clues about their structure and
evolution.  In particular, helioseismology can give us information
about the flows and dynamical structures associated with active regions, about the
depth and extent of thermal disturbances beneath sunspots and other surface phenomena,
and about the nature of the magnetic fields that rise to form active regions.

The frequencies of global oscillations change with the overall level of solar activity 
\citep{WN85,LW90}; this is by 
now very well established \citep[see review by][and references therein]{JCD02}. 
The frequencies of both low degree
modes \citep[e.g.][]{Cetal07} and medium degree modes 
\citep[e.g.][]{Dzetal98,Howeetal99,Dzetal00} increase with solar activity level,
and this frequency increase is stronger at higher frequencies.  Though high degree
global mode measurements are not routinely made, some sets of high degree mode
frequency measurements have been made, and these are found to increase with activity
as well \citep{RSandKorz}.

The use of so-called `ring diagrams' \citep{Hill88} to study solar structure is by now
a well-established technique in helioseismology \citep[see review by][]{GB05}.  In this paper, we
will discuss the changes to mode parameters in ring diagrams between active regions
and the surrounding quiet Sun. Ring diagrams prove useful for studying the effects 
of solar activity on helioseismic modes for two reasons.  First, rings are measured 
on a localized patch of the Sun, providing both temporal and spatial resolution.  
This allows us to isolate phenomena
associated with solar activity from the surrounding Sun.  Second, measurements
can be made to much higher degree much more easily than with global mode analysis.
Ring diagrams were first used by \citet{Hill88} to measure flow rates near the
solar surface.  Rings have been used subsequently by a number of authors to study
the near-surface dynamics of the Sun \citep[e.g.][]{SchouBogart98,BAT99,Haberetal99}. 
Ring diagrams have also been used to study the thermal structure beneath sunspots 
and in the near-surface layers \citep{Basuetal2004,Basuetal07,Bogartetal08}.

The effects of active regions on ring diagram mode parameters have been studied
in previous works.  \citet{Hindetal2000} found that the high degree mode frequencies
obtained from ring diagrams were enhanced in active regions.  \citet{Raj01} found
that the mode widths were also enhanced in active regions, while the amplitudes
were suppressed, and \citet{CRSetal08} found that the relation between the
change in width and mode amplitude was very nearly linear.  \citet{Howeetal04}
assumed that the relation between activity and mode parameter variation was
quadratic, and found a peak increase in mode frequency at around 5mHz, with
a corresponding maximum change in width and amplitude.

In this work, we study the properties of 264 active region ring diagrams
from solar cycle 23.  This sample gives us a statistically significant number
of measurements over a substantial range of activity levels from all phases of
the solar cycle, when compared to earlier works.

\section{Data}
The data are resolved line-of-sight velocity measurements of the solar surface taken by the
Michelson Doppler Imager (MDI) on board the {\it Solar and Heliospheric Observatory\/}
(SOHO) spacecraft \citep{MDI}.  Ring diagrams are constructed from small, $16^\circ \times 16^\circ$ patches 
of the solar disk, tracked to move with the solar differential rotation.  The regions
are tracked across the solar disk center.  The observation cadence is 1 minute, and the
patches are tracked for 8192 minutes.  They are projected onto a rectangular grid
using Postel's projection and corrected for the distortion in the MDI optics.  The Doppler
cubes are then Fourier transformed to obtain a three dimensional power spectrum.  A more
detailed discussion of the construction of ring diagrams can be found in \citet{Patron1997}
and \citet{BAT99}.

The spectra are fit in the same way as \citet{Basuetal2004}.  The functional form is from
\citet{BAT99}; it is an asymmetric Lorentzian profile with terms for advection in both 
transverse directions and for azimuthally asymmetric power distribution. Fits are done 
at constant frequency, so that to interpret these measurements as discrete normal modes, 
we interpolate along each ridge to integer values of the degree $\ell$. Errors are treated 
in the same way as \citep{Basuetal2004}.

Target active regions are selected from the NOAA active region catalog.  Ring diagrams require higher
resolution Dopplergrams than the usual MDI `medium-$\ell$' data that are collected
continuously.  We use full-disk Dopplergram data which are one of MDI's `high rate'
data products, produced mainly during yearly two to three month dynamics run campaigns.
We require a data coverage of greater than 80\%. The NOAA active region catalog includes 
sunspot classifications according to the Mt.~Wilson scheme. The Mt.~Wilson classifications 
are $\alpha$, $\beta$, $\gamma$, and $\delta$, and combinations of these. The NOAA catalog 
lists active regions for each day that they are visible, and in the case where the 
classification changes, we take the classification closest to center disk crossing. 
In our sample, 90\% of the active regions are classified as either $\alpha$, which are 
unipolar groups, or $\beta$, which are simple bipolar groups.

Ring diagrams suffer from a number of systematic effects.  These arise primarily
from foreshortening when away from disk center, and secular changes in the
characteristics of the MDI instrument.  These effects can be minimized by studying
the mode parameters of an active region relative to a quiet Sun region, tracked
at the same latitude and as near in time as possible.  For this sample, two quiet
regions were chosen for each active region, one on either side of the active region.

In order to characterize the level of activity in our rings, we use a measure of
the total unsigned line-of-sight magnetic flux.  This measure is the Magnetic
Activity Index (MAI), which is described in \citet{Basuetal2004} and is a measure of the 
strong magnetic unsigned flux in the ring diagram aperture. In brief, the MAI is 
the spatial and temporal average of all MDI magnetogram pixels with a flux greater 
than 50G, taken over the same aperture and time range as the velocity data for the 
ring diagrams. The activity in each region, then, is characterized by a single number, 
and differential measurements can be characterized by the difference in MAI, 
$\Delta \,\textrm{MAI}$, between the active region and the quiet comparison region. 
We also fit a linear trend in time to the spatial averages of the magnetograms. This 
gives a rough measure of the growth or decay of the active region, and we refer to 
this quantity as the growth parameter.

\section{Analysis}
\begin{figure}
\includegraphics[width=7in]{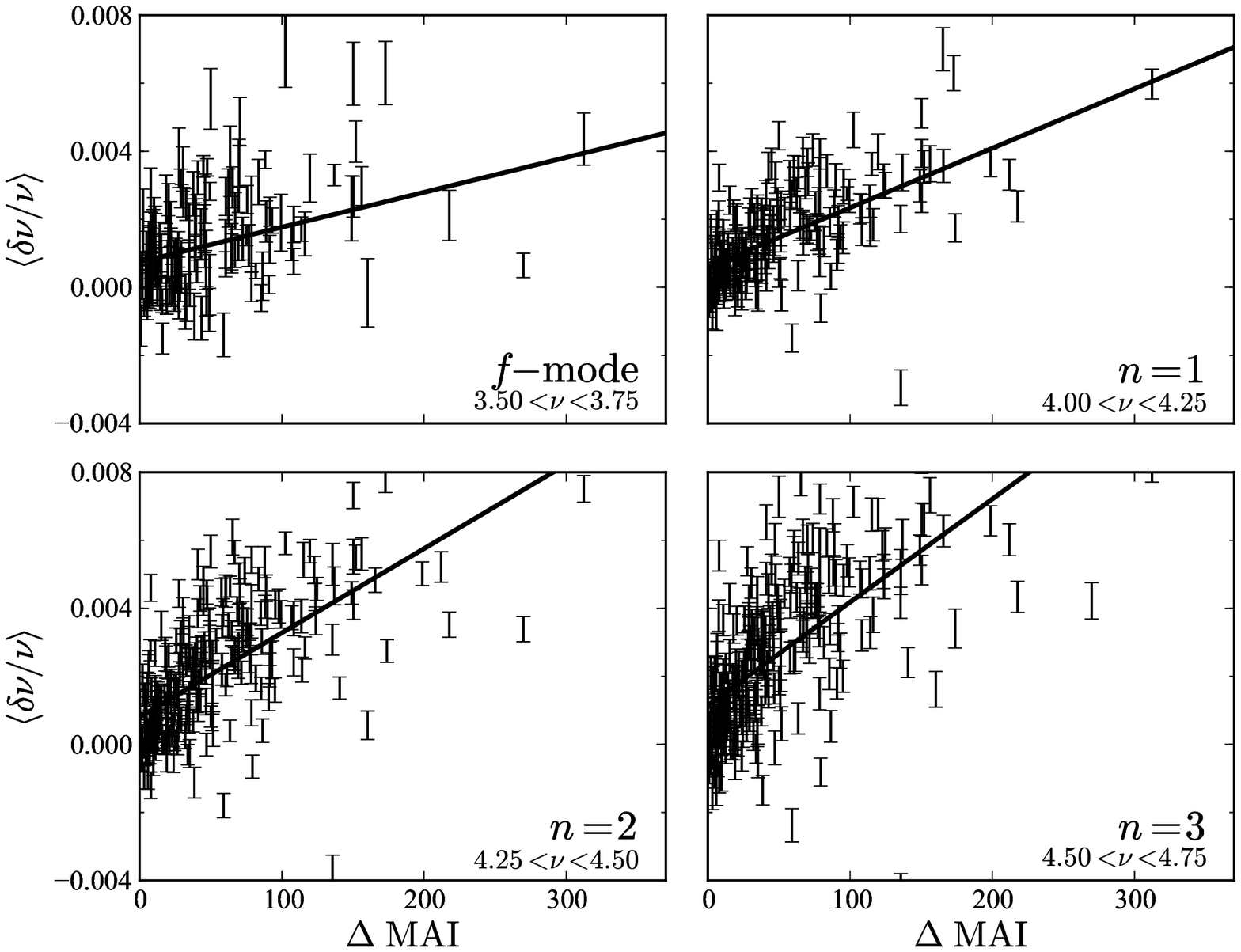}
\caption{Averaged frequency differences are for the $f$-mode and first 
three $p$-modes, as a function of the difference in MAI. The differences 
are with respect to a quiet region near the active region. The range of 
frequencies over which the averages were taken is shown in each panel. 
The straight lines are the best linear fits to the data.
\label{fig:dnu}}
\end{figure}

Mode parameters for all rings and comparison regions were fit using the profile 
from \citet{BAT99}.  We interpolate the ridge fits to integer values of $\ell$,
and take the differences in frequency $\delta \nu$.  The sense of the 
differences are active minus quiet.  In Figure \ref{fig:dnu}, we show averages 
of the scaled frequency differences $\left< \delta \nu / \nu \right>$ for the $f$-mode and 
the first three $p$-modes. The frequency ranges chosen have relatively 
small errors and were fit in most regions in the sample. The behavior seen in 
Figure \ref{fig:dnu} is nevertheless representative of the rest of the data in our 
sample. The majority of $\Delta \,\textrm{MAI}$ values are fairly small --- this is
in part due to the fact that the majority of active regions do not have particularly 
large MAI values, but also due to the fact that, at high activity levels in particular, 
it is sometimes difficult to find a suitable quiet region for comparison. The 
comparison regions therefore sometimes have fairly significant MAI values.

We confirm earlier results finding that, in the presence of magnetic fields,
helioseismic frequencies increase.  We find that, for small and intermediate
magnetic field strengths ($\Delta \,\textrm{MAI}\lesssim 200$~G), the relationship is
linear. For the $p$-mode differences, the correlation coefficients mostly range between
0.6 and 0.7.  The $f$-mode frequencies have larger errors, and the correlation
coefficients are correspondingly lower.  The best fit linear regression lines
are also shown.  The reduced $\chi^2$ values computed from the residuals are 
large --- between 3 and 15 for sets of averages.  Thus, the computed errors
in the fitted frequencies do not completely account for the observed scatter
in the frequency differences. It is possible that a linear relation does not 
completely account for the dependence of frequency on $\Delta \,\textrm{MAI}$, 
but we have tested a variety of higher order polynomials and other functional 
forms without substantially reducing the $\chi^2$ values. When fitting higher
order polynomials, the best fits return essentially the same linear fits as shown
in Figure \ref{fig:dnu}, with negligibly small non-linear coefficients. Other 
functional forms generally increased the $\chi^2$ values.

As a test of the assumption of linearity, we apply an Anderson-Darling test to the 
residuals from the linear best fits. The residuals from the linear regression fit fail
an Anderson-Darling test at the 1\% level. This implies either that the errors are 
non-gaussian, or that the relation between magnetic activity and shift in frequency 
is not entirely linear.

At high magnetic field strengths ($\Delta \,\textrm{MAI}\gtrsim 200$~G), a visual inspection
of Figure \ref{fig:dnu} might suggest a `saturation' effect in the frequency
shifts. Evidence for such an effect in thermodynamic perturbations beneath
active regions has been suggested \citep{Basuetal2004,Bogartetal08}. In this case, however,
we do not find that the outliers at high field strengths are statistically
significant. Fitting different slopes at low and high activity does not change 
the $\chi^2$ values of the residuals, and removing the high activity regions from 
the sample does not substantially change either the $\chi^2$ values or the 
Anderson-Darling $A^2$ statistic of the residuals. Further, we have tested 
whether or not the presence of some comparison regions with significant activity 
is introducing bias. We find that, when the sample is divided between quiet and 
moderately active comparison regions, the frequency differences are indistinguishable. 
It must be noted, however, that a saturation 
effect is not {\it inconsistent\/} with our data, either. A larger sample of high 
field strength regions will be required to adequately address this problem.

\begin{figure}
\includegraphics[width=7in]{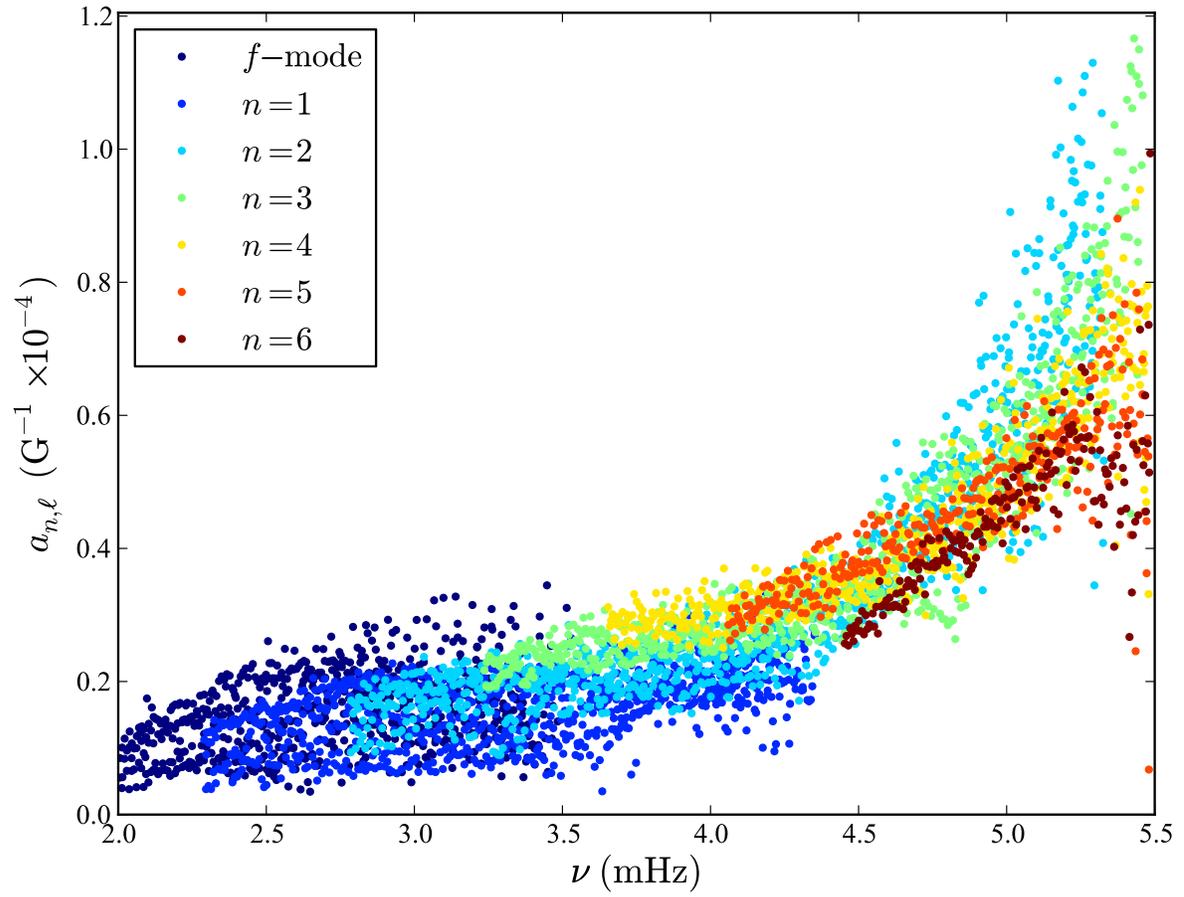}
\caption{Slopes $a_{n,\ell}$ of individual mode frequency shifts with magnetic activity
as a function of frequency up to $\nu=5.5$~mHz. Different colors are used
for different radial orders $n$.
\label{fig:nuslope}}
\end{figure}

To show the behavior of the dependence of frequency on activity, we fit 
a straight line to each individual mode difference as a function of MAI:
\begin{equation}
\frac{\delta \nu_{n,\ell}}{\nu_{n,\ell}} = a_{n,\ell} \Delta\,\textrm{MAI} + b_{n,\ell}.
\end{equation}
The slopes $a_{n,\ell}$
are shown in Figure \ref{fig:nuslope}. The slopes in frequency shifts are
seen to increase with frequency. It is notable, however, that the slopes
are not a pure function of frequency --- different $n$-ridges have slightly
different dependences on frequency. At high frequency, there is some 
indication that the slopes turn over. It has been observed in MDI data 
\citep{Howe08} and in Helioseismic and Magnetic Imager (HMI) data 
\citep{Howe11} that for frequencies above 
the acoustic cutoff, the slopes of frequency with magnetic field strength 
do in fact become negative. In our own sample, however, the quality of the 
fits do not permit us to extend our work much beyond the acoustic cutoff.

\begin{figure}
\includegraphics[width=7in]{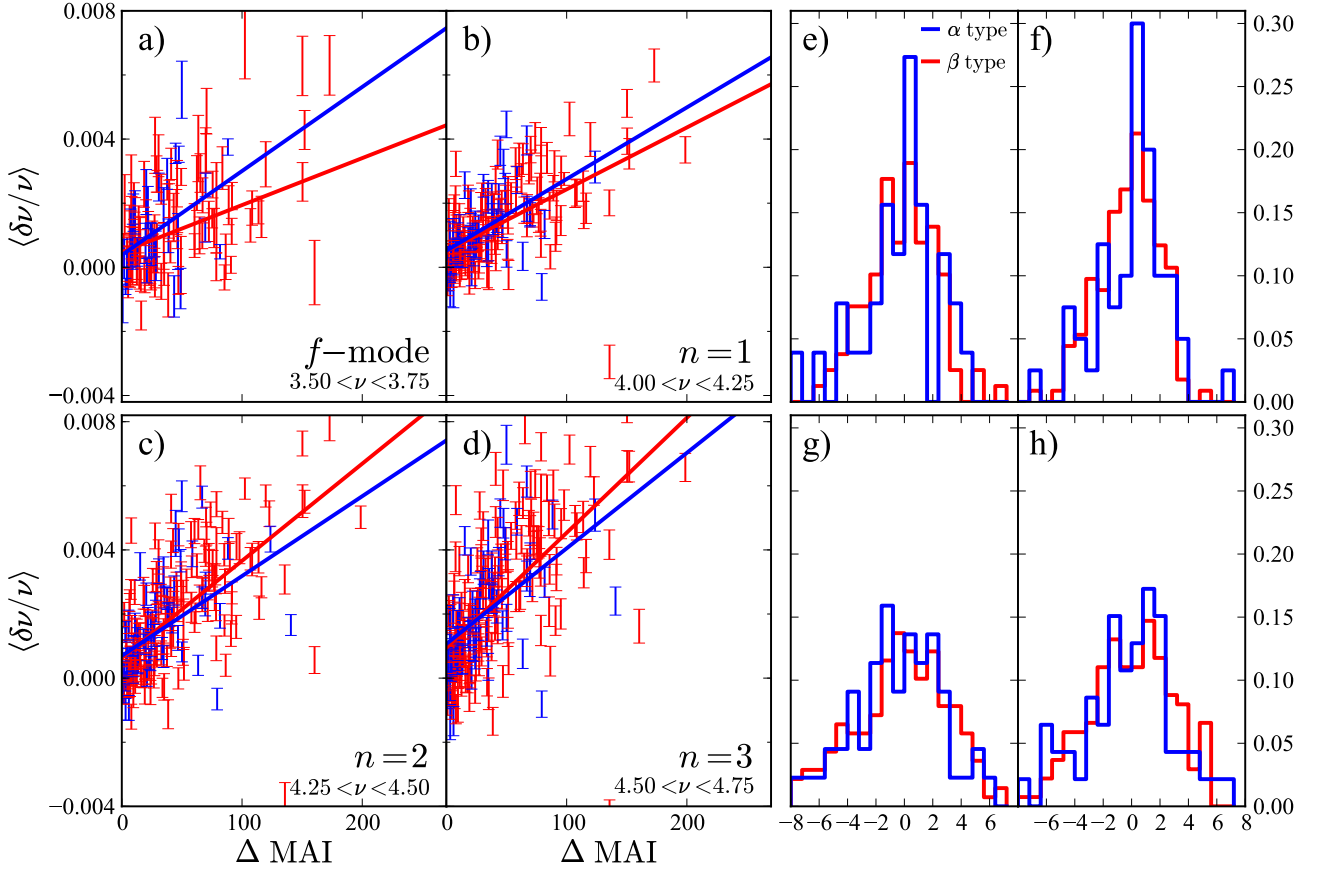}
\caption{Averaged frequency differences separated by spot type for the 
$f$-mode and the first three $p$-modes as a function of the change in 
magnetic activity $\Delta\,\textrm{MAI}$. The frequency ranges are shown 
in panels a) --- d). Blue points denote $\alpha$-type sunspots, red points 
denote $\beta$-type sunspots. Histograms of the residuals are also 
shown in panels e) --- h).
\label{fig:ars}}
\end{figure}

We have examined the differences between different classes of sunspots
in the centers of active regions. The majority of the active regions
in our sample are classified as $\alpha$ or $\beta$ types. When subdivided by
type, the linear correlation is somewhat changed (though the correlation 
coefficients are not), and the distribution
of the residuals for for the $\alpha$-type active regions now satisfy an
Anderson-Darling test for normality at the 15\% level. The residuals for the
$\beta$-type active regions do pass the 10\% test, but do not do as well as
the $\alpha$-types. The slopes of the fits and the correlation coefficients are
not changed by a statistically significant amount. In Figure \ref{fig:ars} we 
show examples of $\left< \delta \nu \right>$ separated by active region type, 
and the residuals to a linear fit.

\begin{figure}
\includegraphics[width=7in]{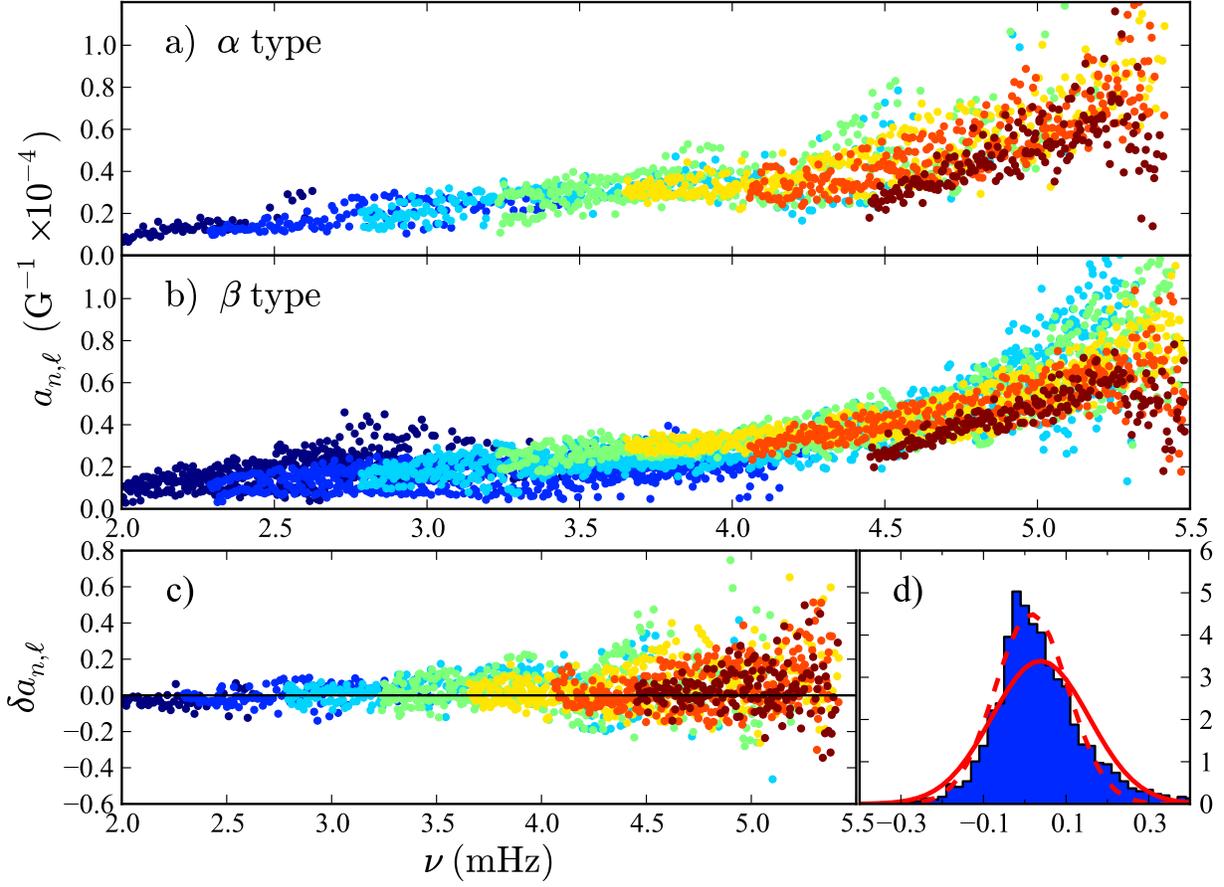}
\caption{Differences in the slopes of individual mode frequency shifts with 
magnetic activity between $\alpha$-type active regions and $\beta$-type active 
regions. Panels a) and b) show the slopes $a_{n,\ell}$ fit to the 
sample of $\alpha$-type spots and $\beta$-type spots, respectively. Panel c) 
shows the differences $\delta a_{n,\ell}$ between the two sets of slopes. 
The differences are shown as a function of frequency up to $\nu=5.5$~mHz. 
The sense of the difference is $\alpha - \beta$. Different colors are used
for different radial orders $n$, and are the same as Figure \ref{fig:nuslope}. 
Panel d) shows a histogram of the differences from panel c). The solid line is 
the normal curve with the mean and standard deviation of the data; the dashed 
line shows the best fit gaussian.
\label{fig:comp}}
\end{figure}

The differences in the slopes between the $\alpha$-type regions and the $\beta$-type 
regions are shown in Figure \ref{fig:comp}. There is a small but systematic trend 
with frequency in the slopes. For $\nu \lesssim 3.5$~mHz, $\alpha$-type spots 
have very slightly smaller slopes $a_{n,\ell}$, while they are slightly larger 
for modes with $\nu > 3.5$~mHz.

We have also subdivided the sample in a number of different ways to attempt to
determine if the mode parameters are sensitive to other variables than the
total magnetic field strength in the region. The fitted parameters were
checked for secular trends with time, or for any dependencies on the latitude
of the region. We have found no statistically significant differences in the
mode parameters or their behavior with increasing activity, either as a
function of time over solar cycle 23 or as a function of latitude. We can
also split the sample in two using the growth parameter of the MAI as a
rough measure of whether or not a given active region is growing or decaying.
The statistical properties of these two sub-samples are indistinguishable from
their parent sample. The confidence level for all of these claims is at the 95\% 
level or greater.

\section{Discussion}
It is by now well established that helioseismic frequencies below the acoustic cutoff 
frequency are enhanced in the presence of magnetic fields. In this work, we have 
measured mode frequencies for a large sample of active regions using ring diagram 
analysis. We find, consistent with most earlier work, that shifts in frequency are 
linearly correlated with the magnetic activity measured at the surface. These 
correlations hold for all mode sets and for the whole range of frequencies studied.

Both \citet{Howeetal04} and \citet{CRSetal08} have presented frequency shifts with 
ring diagrams at various levels of magnetic activity. Our results are generally 
quantitatively consistent with theirs, though we do not find any evidence of a 
quadratic dependence of the mode parameters on activity, as \citet{Howeetal04} 
did. \citet{CRSetal08} find what appears to be 
a much cleaner relation with frequency than we do, in spite of the fact that they 
use the same frequency fitting technique and data from the same instrument. This 
is most likely due to the much smaller number of regions used in that study.

Our finding that the helioseismic frequencies of different active region types 
may be drawn from different statistical populations is novel, but it is not immediately 
clear what the significance of this result is. It implies that the surface geometry, 
not only strength, of the magnetic fields in an active region may determine the oscillation 
characteristics of the Sun.

It has been shown that the inclination of the magnetic fields can affect the amount 
of acoustic power absorption \citep[e.g.][]{Callyetal03}. Observational evidence for 
this has been found using helioseismic holography \citep{Schunkeretal05,Schunkeretal08} 
and time-distance analysis \citep{Z&K06}. It is possible that we are seeing this 
effect averaged out over the much larger region sampled in ring diagram analysis. 

Although we do find that the slopes $a_{n,\ell}$ are slightly different for the 
two different active region types, the fact that these differences appear to be 
simply a function of frequency implies that it is very unlikely do be due to 
changes in the thermal structure in the regions typically resolved in ring 
diagram structure inversions. In the region where the upper turning points of 
the modes are located, however, it is possible that a different thermal 
stratification in different sunspot types would give rise to a difference 
such as the one we have detected, since upper turning points are shallower 
with increasing frequency. The direct effects of magnetic fields also have 
an effect on the upper turning points, as was shown by \citet{Jain07} in 
the case of horizontal magnetic fields.

The errors in the measurement of the MAIs are small, but it is possible that 
small systematic errors arise from the fact that the calculation is restricted to 
the line-of-sight fields. Different surface field configurations could give rise 
to somewhat different systematic errors. It is difficult to estimate how large 
this effect could be at present, but continuously available high-resolution data 
becoming available from the Helioseismic and Magnetic Imager (HMI) on the Solar 
Dynamics Observatory (SDO) is allowing full vector inversions for magnetic fields. 
This will allow us to more accurately compute the total magnetic fields in active 
regions.

\acknowledgements
CSB is supported by a NASA Earth and Space Sciences fellowship NNX08AY41H. 
SB is supported by NSF grants ATM 0348837 and ATM 0737770 to SB. 
This work utilizes data from the Solar Oscillations Investigation/
Michelson Doppler Imager (SOI/MDI) on the Solar and Heliospheric
Observatory (SOHO).  SOHO is a project of international cooperation
between ESA and NASA.  MDI is supported by NASA grants NAG5-8878
and NAG5-10483 to Stanford University.

\begin{thebibliography}

\bibitem[Basu et al.(2004)]{Basuetal2004} Basu, S., Antia, H.~M., \& Bogart, R.~S.\ 2004, \apj, 610, 1157

\bibitem[Basu et al.(2007)]{Basuetal07} Basu, S., Antia, H.~M., 
\& Bogart, R.~S.\ 2007, \apj, 654, 1146

\bibitem[Basu et al.(1999)]{BAT99} Basu, S., Antia, H.~M., \& Tripathy, S.~C.\ 1999, \apj, 512, 458

\bibitem[Bogart et al.(2008)]{Bogartetal08} Bogart, R.~S., Basu, S., 
Rabello-Soares, M.~C., \& Antia, H.~M.\ 2008, \solphys, 251, 439

\bibitem[Cally et al.(2003)]{Callyetal03} Cally, P.~S., Crouch, 
A.~D., \& Braun, D.~C.\ 2003, \mnras, 346, 381

\bibitem[Chaplin et al.(2007)]{Cetal07} Chaplin, W.~J., 
Elsworth, Y., Miller, B.~A., Verner, G.~A., 
\& New, R.\ 2007, \apj, 659, 1749

\bibitem[Christensen-Dalsgaard(2002)]{JCD02}
Christensen-Dalsgaard, J.\ 2002, Reviews of Modern Physics, 74, 1073

\bibitem[Dziembowski et al.(1998)]{Dzetal98} Dziembowski, W.~A., 
Goode, P.~R., di Mauro, M.~P., Kosovichev, A.~G., 
\& Schou, J.\ 1998, \apj, 509, 456

\bibitem[Dziembowski et al.(2000)]{Dzetal00} Dziembowski, W.~A., 
Goode, P.~R., Kosovichev, A.~G., \& Schou, J.\ 2000, \apj, 537, 1026

\bibitem[Gizon 
\& Birch(2005)]{GB05} Gizon, L., \& Birch, A.~C.\ 2005, Living Reviews in Solar Physics, 2, 6

\bibitem[Haber et al.(1999)]{Haberetal99} Haber, D.~A., Hindman, B.~W., Toomre, J., Bogart, R.~S., Schou, J., 
\& Hill, F.\ 1999, SOHO-9 Workshop on Helioseismic Diagnostics of Solar Convection and Activity, 9,

\bibitem[Hindman et al.(2000)]{Hindetal2000} Hindman, B., Haber, D., Toomre, J., \& Bogart, R.\ 2000, \solphys, 192, 363

\bibitem[Hill (1988)]{Hill88} Hill, F.\ 1988, \apj, 333, 996

\bibitem[Howe et al.(1999)]{Howeetal99} Howe, R., Komm, R., 
\& Hill, F.\ 1999, \apj, 524, 1084

\bibitem[Howe et al.(2004)]{Howeetal04} Howe, R., Komm, R.~W., Hill, F., Haber, D.~A., \& Hindman, B.~W.\ 2004, \apj, 608, 562

\bibitem[Howe et al.(2008)]{Howe08} Howe, R., Haber, D.~A., 
Hindman, B.~W., Komm, R., Hill, F., 
\& Gonzalez Hernandez, I.\ 2008, Subsurface and Atmospheric Influences on Solar Activity, 383, 305

\bibitem[Howe et al.(2011)]{Howe11} Howe, R., Tripathy, S., 
Gonz{\'a}lez Hern{\'a}ndez, I., Komm, R., Hill, F., Bogart, R., 
\& Haber, D.\ 2011, Journal of Physics Conference Series, 271, 012015

\bibitem[Jain(2007)]{Jain07} Jain, R.\ 2007, \apj, 656, 610

\bibitem[Libbrecht \& Woodard(1990)]{LW90} Libbrecht,
K.~G., \& Woodard, M.~F.\ 1990, \nat, 345, 779

\bibitem[Patron et al.(1997)]{Patron1997} Patron, J., et al.\ 1997, \apj, 485, 869

\bibitem[Rabello-Soares \& Korzennik(2009)]{RSandKorz} Rabello-Soares, M.~C., \& Korzennik, S.~G.\ 2009, Astronomical Society of the Pacific Conference Series, 416, 277

\bibitem[Rabello-Soares et al.(2008)]{CRSetal08} Rabello-Soares, M.~C., Bogart, R.~S., \& Basu, S.\ 2008, Journal of Physics Conference Series, 118, 012084

\bibitem[Rajaguru et al.(2001)]{Raj01} Rajaguru, S.~P., Basu, S., \& Antia, H.~M.\ 2001, \apj, 563, 410

\bibitem[Scherrer et al.(1995)]{MDI} Scherrer, P.~H., et 
al.\ 1995, \solphys, 162, 129

\bibitem[Schou \& Bogart(1998)]{SchouBogart98} Schou, J., \& Bogart, R.~S.\ 1998, \apjl, 504, L131

\bibitem[Schunker et al.(2005)]{Schunkeretal05} Schunker, H., Braun, 
D.~C., Cally, P.~S., \& Lindsey, C.\ 2005, \apjl, 621, L149

\bibitem[Schunker et al.(2008)]{Schunkeretal08} Schunker, H., Braun, 
D.~C., Lindsey, C., \& Cally, P.~S.\ 2008, \solphys, 251, 341

\bibitem[Woodard \& Noyes(1985)]{WN85} Woodard, M.~F.,
\& Noyes, R.~W.\ 1985, \nat, 318, 449

\bibitem[Zhao 
\& Kosovichev(2006)]{Z&K06} Zhao, J., \& Kosovichev, A.~G.\ 2006, \apj, 643, 1317

\end{thebibliography}
\end{document}